\documentclass{aa}
\usepackage{txfonts}
\usepackage[version=4]{mhchem}
\usepackage{natbib}
\bibpunct{(}{)}{;}{a}{}{,}
\usepackage{graphicx}
\usepackage{multicol}
\usepackage[dvipsnames]{xcolor}
\usepackage{geometry}
\nopagebreak[0]
\begin{document} 

   \title{Infrared action spectroscopy of doubly charged PAHs and their contribution to the aromatic infrared bands}
   
   \author{S. Banhatti\inst{1}, J. Palot\'as\inst{2}, P. Jusko\inst{3}, B. Redlich\inst{2} , J. Oomens\inst{2,4}, S. Schlemmer\inst{1},  and S. Brünken\inst{2} 
          }

   \institute{I. Physikalisches Institut, Universität zu Köln, Zülpicher Str. 77, 50937 Köln, Germany\\
\email{banhatti@ph1.uni-koeln.de}
         \and
             Radboud University, Institute for Molecules and Materials, FELIX Laboratory, Toernooiveld 7, 6525ED Nijmegen, the Netherlands\\
             \email{sandra.bruenken@ru.nl}
        \and
            Max Planck Institute for Extraterrestrial Physics, Gießenbachstraße 1, 85748 Garching, Germany
        \and
            van ’t Hoff Institute for Molecular Sciences, University of Amsterdam, Science Park 908, 1098XH Amsterdam, the Netherlands
            }
   \date{15 February 2020 ; accepted manuscript }

\abstract
   {The so-called aromatic infrared bands (AIBs) are attributed to emission of polycyclic aromatic hydrocarbons (PAHs). The observed variations toward different regions in space are believed to be caused by contributions of different classes of PAH molecules, that is to say with respect to their size, structure, and charge state. Laboratory spectra of members of these classes are needed to compare them to observations  and to benchmark quantum-chemically computed spectra of these species. In this paper we present the experimental infrared (IR) spectra of three different PAH dications, naphthalene$^{2+}$, anthracene$^{2+}$, and phenanthrene$^{2+}$, in the vibrational fingerprint region 500 - 1700~cm$^{-1}$. The dications were produced by electron impact ionization (EI) of the vapors with 70\,eV electrons, and they remained stable against dissociation and Coulomb explosion. The vibrational spectra were obtained by IR predissociation of the PAH$^{2+}$ complexed with neon in a 22-pole cryogenic ion trap setup coupled to a free-electron infrared laser at the Free-Electron Lasers for Infrared eXperiments (FELIX) Laboratory.  We performed anharmonic density-functional theory (DFT) calculations for both singly and doubly charged states of the three molecules. The experimental band positions showed excellent agreement with the calculated band positions of the singlet electronic ground state for all three doubly charged species, indicating its higher stability over the triplet state. The presence of several strong combination bands and additional weaker features in the recorded spectra, especially in the 10-15~$\mu$m region of the mid-IR spectrum, required anharmonic calculations to understand their effects on the total integrated intensity for the different charge states. These measurements, in tandem with theoretical calculations, will help in the identification of this specific class of doubly-charged PAHs as carriers of AIBs. 
}


\maketitle

\section{Introduction}
The widespread mid-infrared (mid-IR) emission features observed in many astrophysical objects, such as HII regions, planetary nebulae (PNe), reflection nebulae, and young stellar objects, have been a topic of great interest  since their detection in the 1970s and 1980s. Strong bands are observed at 3.3, 6.2, 7.7, 11.3, and 12.7 $\mu$m along with several weak features, which together make up the unidentified infrared (UIR) or aromatic infrared bands (AIBs) \citep{Sellgren1984,Sellgren1985,Jourdain1990,Cohen1986}. These bands are hypothesized to arise from a family of polycyclic aromatic hydrocarbons (PAHs) excited by absorption of ultraviolet (UV) radiation from nearby stars and their subsequent emission in the mid-IR region \citep{LP1984,ATB1989,Allamandola1985,Hudgins1997}

Depending on the conditions in the interstellar medium (ISM), different families of PAH molecules are proposed to exist, such as neutral and ionic variants of different size, shape, and  hydrogenation states, all of which have an effect on the relative band intensities and band positions in the observed UIR bands \citep{Hony2001, Dartois1997}. For example, the relative intensity variation between the {6.2, 7.7, 8.6, and 11.2 $\mu$m} bands has been attributed to the degree of ionization of PAHs \citep{Galliano2008}. Also, the 18.9 $\mu$m band has been identified as a signature of multiply charged PAHs \citep{Tielens2008}.  

Identifying different classes of PAHs as carriers of UIR bands requires laboratory data of their gas-phase  IR spectra. Several schemes have been used over the past decades to record the IR spectra of PAHs. In particular for the study of charged PAHs, the development of intense and widely tunable free-electron lasers in facilities such as the Free-Electron Lasers for Infrared eXperiments (FELIX) Laboratory \citep{OMA1995}\footnote{\url{https://www.ru.nl/felix/}} and CLIO\footnote{\url{http://old.clio.lcp.u-psud.fr/clio_eng/clio_eng.htm}} has made it possible to implement sensitive action spectroscopic schemes such as infrared multiple photon dissociation (IRMPD). Gas-phase IR spectra of several small- to medium-sized cationic species  have been recorded with the IRMPD scheme (e.g., naphthalene, phenanthrene, anthracene, coronene, and protonated naphthalene), pioneered by Oomens and coworkers \citep{Oomens2000,Oomens2001,BRV2011,Dofer2007}. More recent works on the IR spectroscopy of gas phase PAH ions include larger species such as  diindenoperylene (\ce{C32H16$^+$}), dicoronylene (\ce{C48H20$^+$}), hexa-peri-hexabenzocoronene (\ce{C42H18$^+$}) \citep{Zhen2017,Zhen2018}, and rubicene  (\ce{C26H14$^+$}) \citep{Bouwman2020} and PAH anions of naphthyl, anthracenyl, and pyrenyl \citep{gao2014}. Their spectra have been shown to contain vibrational modes which could account for some of the features observed in the UIR bands. Pentagon-containing PAHs, for example, were shown to possess distinct vibrational modes at around 1100~cm$^{-1}$ ($\sim 9.3$~$\mu$m), and the ratio of the $9.3/6.2$~$\mu$m band strength was used to estimate their relative abundance in several astronomical sources \citep{Bouwman2020}. Experimental results also verified that protonated nitrogen-containing species (H$^+$PANHs) might indeed be responsible for the 6.2~$\mu$m UIR emission \citep{GPO2010}. 

Another powerful action spectroscopic technique is infrared predissociation (IRPD), which requires cold ion conditions to tag the ions with a weakly bound rare-gas atom. Earlier IRPD studies applied to PAH ions used a molecular beam cooled by supersonic expansion, and they provided gas phase spectra of cold protonated \citep{Ricks2009} and cationic \citep{Piest1999} naphthalene and phenanthrene \citep{Piest2001} via dissociation of their weakly bound complexes with Ar. With the advent of cryogenic ion trapping techniques, IRPD spectra can now be recorded at temperatures as low as 4\,K, allowing for the use of weaker bound rare-gas atoms such as He or Ne as tagging agents, introducing smaller shifts of vibrational bands in the experimental spectra \citep{ABK2002,Jasik2014,GNM2017,Gerlich2018,Jusko2019}. The advantage of IRPD compared to multiphoton IRMPD spectroscopy is that dissociation happens after absorption of a single photon, with approximately constant efficiency, resulting in experimental vibrational spectra more closely resembling the linear absorption spectrum of the ions, both in intensity and band positions. In the past, we have successfully used this technique in our cryogenic ion trap instrument coupled to the widely tunable free electron lasers at the FELIX Laboratory to record narrow-linewidth spectra of PAH cations and related species \citep{Jusko2018a,Jusko2018b,Panchagnula2020}, and here we  apply it for the spectroscopic characterization of PAH dications. 

The presence of PAH dications and their formation in the ISM has been discussed previously by \cite{Leach1986}. Theoretical investigations on their stability and vibrational spectra have been carried out by \cite{Malloci2007}, \cite{Bakes2001b,Bakes2001a}, and \cite{Bauschlicher1997}. However, most of the laboratory work done so far in regards to the vibrational spectra of PAHs has been focused on their neutrals and monocations. There is no spectroscopic laboratory data on doubly charged PAHs in the IR regime except for an IRMPD study of hexa-peri-hexabenzocoronene, \ce{C42H18$^{2+}$} \citep{Zhen2017}. In this work we present the IR predissociation spectra of the following three doubly charged PAHs using Ne-tagging in a cryogenic ion trap: naphthalene$^{2+}$ (naph$^{2+}$, \ce{C10H8}$^{2+}$), anthracene$^{2+}$ (anth$^{2+}$), and its isomer phenanthrene$^{2+}$ (phen$^{2+}$, both  \ce{C14H10}$^{2+}$).

\section{Methods}
\subsection{Experimental methods}
\begin{figure}
\centering
\includegraphics[width=8.8cm]{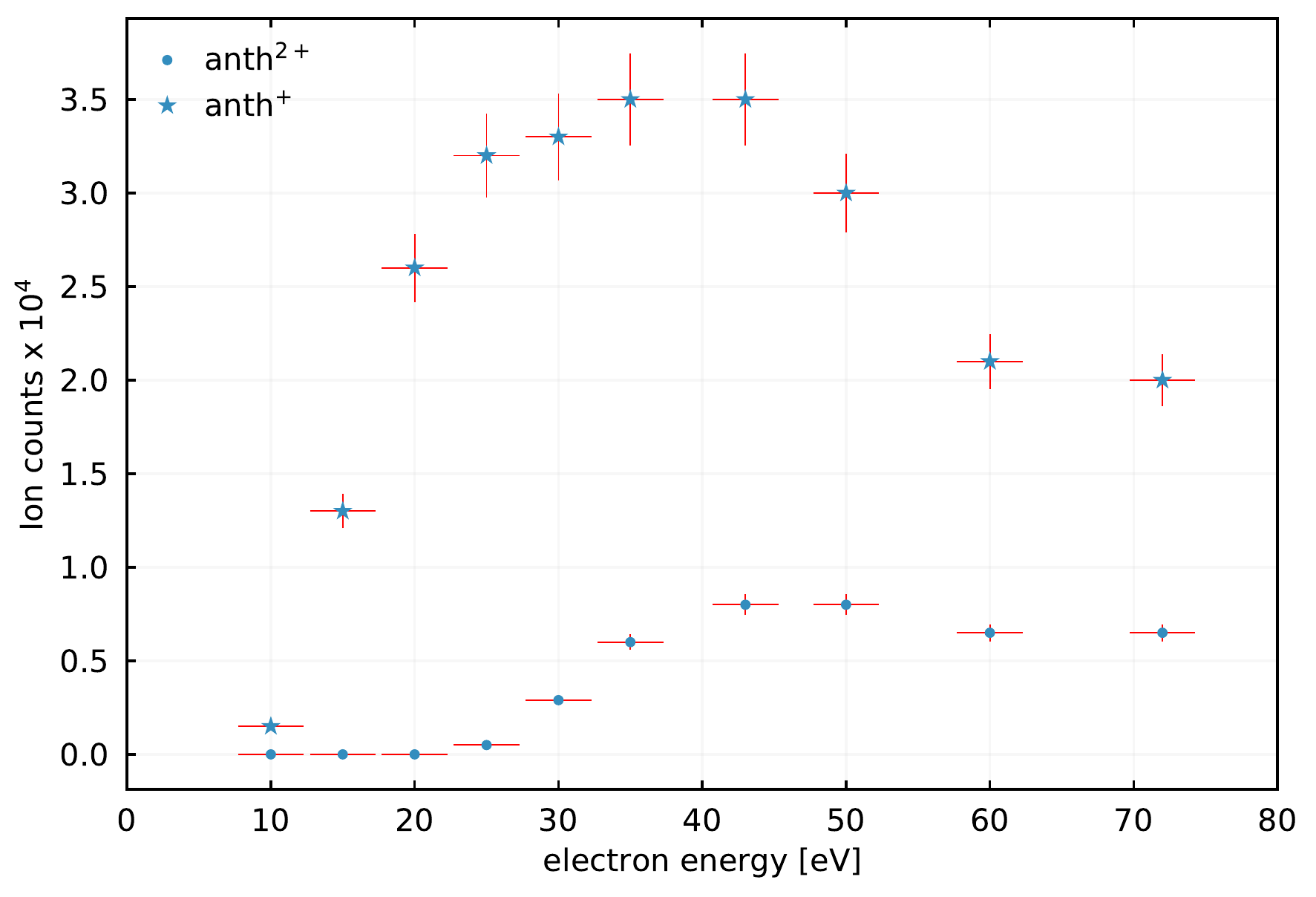}
  \caption{Ionization yield for m/z=89 (anth$^{2+}$) and m/z=178 (anth$^{+}$) produced in a non-storage electron ionization source.}
     \label{ionyield}
\end{figure}

\begin{figure}
\centering
\includegraphics[width=8.8cm]{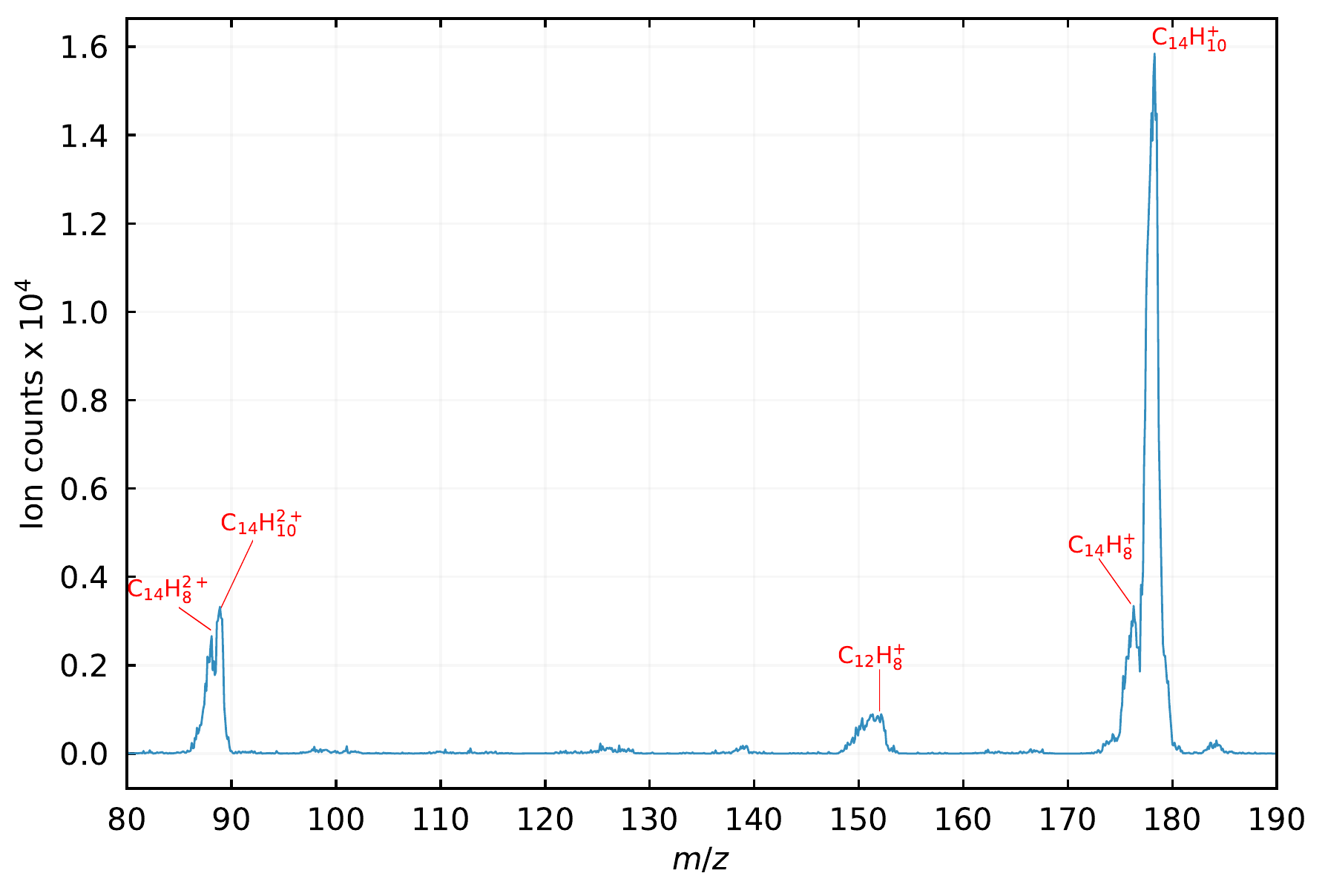}
  \caption{Mass spectrum using phenanthrene precursor in the mass range m/z=80-190. The monocation at m/z=178, the doubly charged parent at m/z=89, and the \ce{C2H2} loss fragment at m/z=152 are shown in the figure. 
          }
     \label{massspec}
\end{figure}

The vibrational spectra of PAH dications were recorded using the FELion cryogenic 22-pole ion trap setup coupled to the free-electron laser FEL-2 at the FELIX Laboratory \citep{OMA1995}. The FELion experimental setup has already been described in detail in \citet{Jusko2019}. To provide a brief overview, the respective vapors of PAHs are ionized by electron impact ionization (EI), which results in the formation of several charged fragments including the doubly charged naphthalene cation (\ce{C10H8^2+}, mass-to-charge ratio of m/z=64, naph$^{2+}$ in the following) and the doubly charged phenanthrene and anthracene cations (\ce{C14H10^2+}, m/z=89, phen$^{2+}$, and anth$^{2+}$, resp.). In our experiments, we used electron energies of 50-70~$\mathrm{eV}$, which is well above the appearance energy of naph$^{2+}$ at 21.52~$\mathrm{eV}$, and phen$^{2+}$  and anth$^{2+}$ around 20~$\mathrm{eV}$ \citep{Holm2011,Malloci2007b,vanderBurgt2018,vanderBurgt2019}. The dication yield was found to be the highest around 50 $\mathrm{eV}$ and remained constant at higher energies, as can be seen in Fig. \ref{ionyield}, showing the anth$^{2+}$ (and anth$^{+}$) yield as a function of electron energy. Fig. \ref{massspec} shows a typical mass spectrum upon EI of phenanthrene vapor. 

While naph$^{2+}$ was produced efficiently with an RF storage source \citep{Gerlich1992}, we could not produce high yields of phen$^{2+}$ or anth$^{2+}$ ions in this way. This is probably due to the chemical quenching of initially produced dications by charge transfer and reactions with the neutral precursor and fragments in the high pressure (10$^{-5}$~mbar) storage source over the typical storage time of seconds. Hence, a non-storage EI source was used for these dications. The ions produced in the source were mass selected to within m/z = $\pm$2 for the desired dication in a quadrupole mass filter. The RF of the quadrupole was tuned such that at least 80\% of the ions transmitted belong to the dication \ce{C14H10}$^{2+}$ or \ce{C10H8}$^{2+}$ (see fig \ref{mass_spec_Ne_tagging}). The mass selected ions were then complexed in situ with neon (Ne) in a 22-pole cryogenic ion trap. Either pure Ne gas (in the case of anth$^{2+}$ and phen$^{2+}$) or a 3:1 He:Ne gas mixture (in the case of naph$^{2+}$) was pulsed into the ion trap for 100-150\,ms at the beginning of the storage period, with the trap held at a temperature of 15\,K or 6.3\,K, respectively. A second quadrupole mass filter with a mass resolution of better than m/z = 0.5 was used to mass select the products formed in the ion trap, thus ensuring that only the target dications complexed with Ne were detected to record the spectrum.   

For a doubly charged ion, the Ne-dication complex appears at a mass-to-charge ratio of  m/z = +10 higher than the PAH$^{2+}$ in the mass spectra, which was observed for all of the abovementioned ions; the tagging yield was of the order 30\,\%. In addition, we also observed doubly and triply tagged ions at m/z = +20 and +30 (see Fig. \ref{mass_spec_Ne_tagging} in the appendix). The Ne-dication complexes are stored in the trap for typically 2.6 seconds where they are irradiated by IR radiation from the free-electron laser FEL-2. The FEL was set to a narrow bandwidth (FWHM) of 0.4-0.5$~\mathrm{\%}$ at 10\,Hz repetition rate and the spectrum was recorded in the range of $500-1800$\,cm$^{-1}$ ($20-5.5 \mu$m) with a typical laser pulse energy inside the trap of $10-40$\,mJ. The IRPD spectra were recorded by measuring the depletion of Ne-dication complex ion counts as a function of laser frequency. To account for fluctuations in the number of complex ions and varying laser energy during the course of a single scan, the spectra were normalized to laser pulse energy and baseline corrected before averaging over multiple scans. Saturation depletion measurements allowed us to decode the isomer ratios of molecular ions having two or more stable structures, see, for example, \cite{Jusko2018a}. A detailed account of this technique is discussed in \cite{Jusko2019}.

\subsection{Theoretical methods}
\label{theo}
Geometry optimizations and frequency calculations were performed at the density functional theory (DFT) level. The Gaussian16 software package \citep{g16} was used for all calculations as installed at the Cartesius supercomputer at SurfSARA, Amsterdam. The vibrational frequencies of doubly charged PAHs were calculated within the harmonic approximation using the hybrid B3LYP functional with 6-311+G(d,p) basis set and scaled uniformly with 0.9679 \citep{Andersson2005}. Furthermore, anharmonic calculations within the vibrational second-order perturbation level of theory (VPT2) were performed as implemented in Gaussian 16 with the same functional and basis set combination. We should note here that the VPT2 method implemented in Gaussian16 only provides a basic treatment of resonances. A more accurate calculation, which is beyond the scope of the present work, would require an explicit treatment of resonating polyads with variational approaches, as demonstrated earlier on the example of PAH molecules \citep{Mackie2015, Mackie2018, Chen2018a, Matteo2015}.

To investigate the effect of Ne-tagging on the vibrational spectra, additional harmonic calculations were performed on the naph$^{2+}$-Ne complexes. Several binding geometries of the Ne atom to the PAH are conceivable with the Ne atom on top of the molecular plane and in plane with the molecule. The five lowest energy conformers were optimized and harmonic vibrational frequencies were calculated (see appendix \ref{Ne_tagging}). Here the dispersion corrected wB97XD functional with a cc-pVTZ basis was used, which was found previously to work well for weakly-bound RG-ion complexes \citep{Jusko2018a}.

\section{Results and discussion}
\subsection{Naphthalene} 
\begin{figure}
    \centering
    \includegraphics[width=8.8cm]{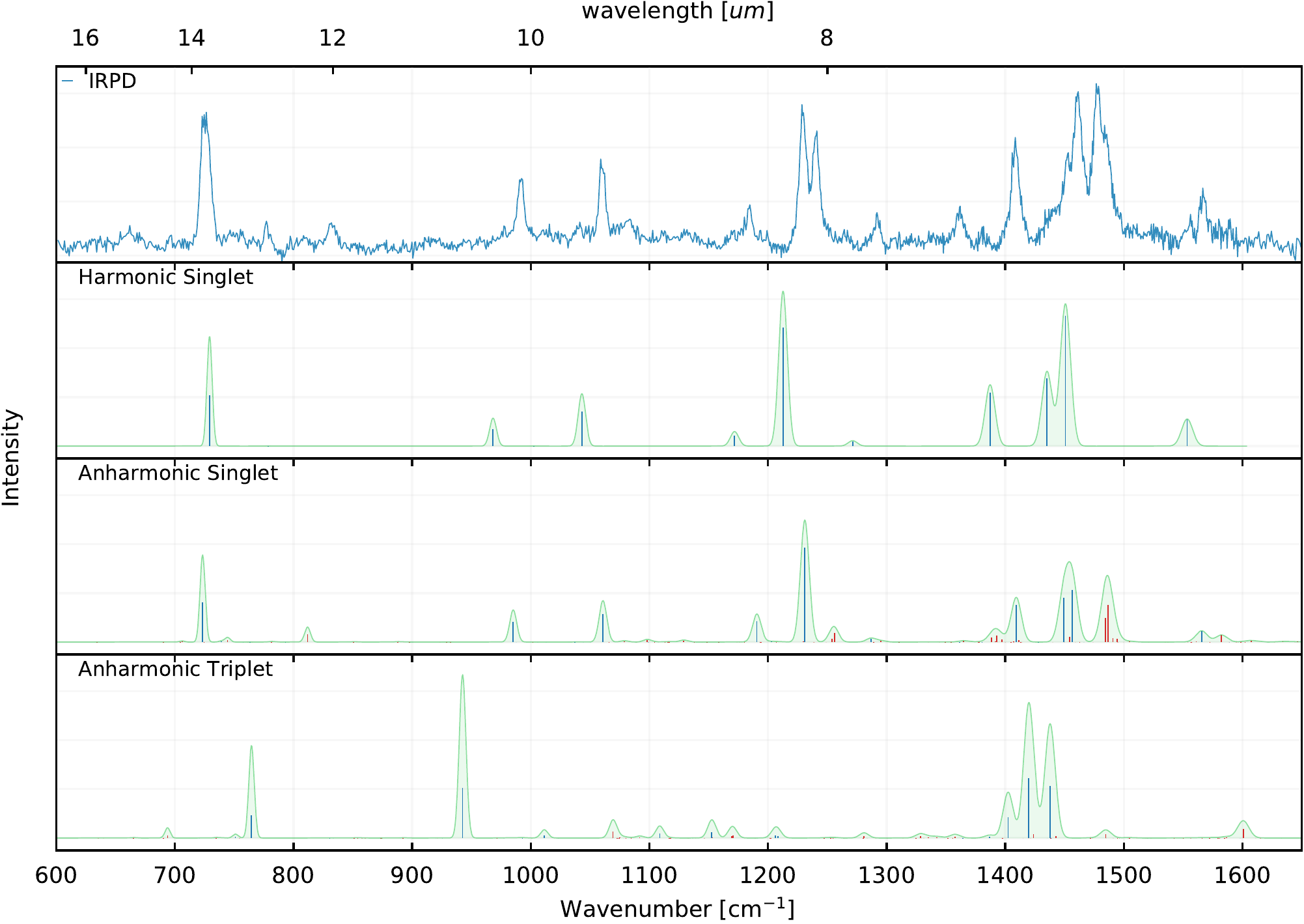} 
    \caption{IRPD spectrum of Ne-tagged naph$^{2+}$ (top panel, blue) compared with calculated anharmonic and harmonic band positions of singlet naph$^{2+}$ and anharmonic frequencies of triplet naph$^{2+}$. Both fundamental modes (blue) and combination modes (red) are shown. The calculated spectrum was convoluted with a Gaussian lineshape function, where the width is given by the FEL bandwidth, and the area corresponds to the calculated intensity in $\mathrm{km\hspace{1pt}mol^{-1}}$.}
    \label{naph_exp_calc_IRMPD}
\end{figure}

The experimental IRPD spectrum of Ne-tagged naph$^{2+}$ (Ne-IRPD) is shown in the top panel of Fig.  \ref{naph_exp_calc_IRMPD}. For comparison, the calculated spectra of the singlet electronic ground state with $^1A_g$ symmetry  and of the 0.31~eV higher lying triplet $^{3}B_{3g}$ electronic state are shown in the panels below. The peak positions from the fitted experimental spectrum and the corresponding band assignments based on the calculations are listed in Table \ref{Naph_table}. Most of the band positions in the Ne-IRPD spectrum can be readily assigned to the singlet state of naph$^{2+}$, and they are in excellent agreement with the calculated anharmonic frequencies to within 10~$\mathrm{cm^{-1}}$ for the fundamental bands.

To account for the influence of the Ne-tag on the vibrational band positions, we compared  harmonic DFT calculations for bare and different conformers of  Ne-tagged naph$^{2+}$. They revealed negligible band shifts of $<$ 1~$\mathrm{cm^{-1}}$ for most of the bands, with only a few bands maximally shifted by 5~$\mathrm{cm^{-1}}$ (see Appendix {\ref{Ne_tagging}}). The IRPD spectrum of the Ne-tagged species can therefore be viewed as an excellent proxy for that of the bare ion.

Only three of the experimental  bands, at 1361~$\mathrm{cm^{-1}}$, 1240~$\mathrm{cm^{-1}}$, and 777~$\mathrm{cm^{-1}}$, could not be assigned in a straightforward manner. The band at  1362~$\mathrm{cm^{-1}}$ lies close to weak combination bands predicted at 1361 and 1365~$\mathrm{cm^{-1}}$ (not listed in Table \ref{Naph_table}). 
Several strong predicted combination bands were assigned to a weak experimental feature at 1381~$\mathrm{cm^{-1}}$, and they could also be assigned to the 1362~$\mathrm{cm^{-1}}$ band if we assume large anharmonic shifts unaccounted for by the calculations. Another feature at 832~$\mathrm{cm^{-1}}$ assigned to a predicted band at 812~$\mathrm{cm^{-1}}$ shows a similarly large deviation. For combination bands, larger deviations of calculated anharmonic band positions are expected when using VPT2, so we tentatively assigned these features to one or a combination of the above combination bands.

The rather strong feature at 1240~$\mathrm{cm^{-1}}$ is close to the strong \ce{C}-\ce{H} in plane bending vibrational mode at 1229~$\mathrm{cm^{-1}}$, which almost appears as a doublet. From the calculations, we can exclude that the symmetry breaking induced by the Ne-tag has a significant influence on the spectrum, and that the presence of multiple Ne-ion isomers are responsible for the observed splitting of the band. We therefore assume that the blue-shifted feature at 1240~$\mathrm{cm^{-1}}$ is due to a combination band of the strong \ce{C}-\ce{H} bending mode with a vibration involving only the Ne-tag, a phenomenon often seen in rare-gas tagging experiments \citep{Bruenken2019}.  The lowest harmonic fundamental frequencies involving the Ne-atom are shown in Fig. \ref{Ne_tagging}. They fall in the $5-60$~cm$^{-1}$ wavenumber range, and they are thus lying close to the observed difference between the two experimental features. However, we should note that our calculations also show a Fermi resonance of the 1229~cm$^{-1}$ mode with a close-lying combination band of the same symmetry, which might gain intensity due to this interaction.
  
The weak band at 777~$\mathrm{cm^{-1}}$ appears to be coincident with a predicted fundamental of triplet naph$^{2+}$ at 764~$\mathrm{cm^{-1}}$, but the absence of a much stronger predicted triplet mode around 943~$\mathrm{cm^{-1}}$ in the experimental spectrum suggests otherwise. To test the possible presence of the electronically excited triplet state, we performed saturation depletion measurements on several strong bands assigned to singlet naph$^{2+}$, revealing an abundance of at least 80 \% of the ground electronic state. This indicates that up to 20\,\% of the formed dications could be in the electronically excited triplet state or a different isomer \citep{Leach1989a,SM2015}. Since we do not observe any other triplet bands as expected from the calculated spectrum in Fig. \ref{naph_exp_calc_IRMPD} (bottom panel), the latter is more likely.
  
Whereas the calculated, scaled harmonic band positions of the singlet state correlate well with those from the anharmonic calculation, we can observe large shifts (up to 20~$\mathrm{cm^{-1}}$) between both levels of theory, indicating the need to include mode dependent anharmonic effects to correctly describe the vibrational PAH dication spectra. Furthermore, the strong combination bands between 1480-1490 ~$\mathrm{cm^{-1}}$ observed in the experimental spectrum can be accounted for only with anharmonic calculations. The narrow linewidths observed in our Ne-IRPD spectrum can serve as a benchmark for anharmonic calculations of PAH dications just as previously shown for PAH neutrals \citep{Mackie2015,Maltseva2015}.

\subsection{Phenanthrene}
The IRPD spectrum of phen$^{2+}-$Ne compares well with the calculated anharmonic spectrum of the $^1A_1$ singlet electronic state as seen in Fig. \ref{All_spec}. Table \ref{Phen_table} shows the peak positions with the assigned anharmonic and harmonic frequencies with many of the peak positions coinciding within 10~$\mathrm{cm^{-1}}$ of the calculated anharmonic spectrum for the singlet ground state. Again, we do not observe any bands that could be assigned to the triplet $^3B_2$ electronic state 0.44~eV higher (\ref{Triplet_table}).

The bands below 800~$\mathrm{cm^{-1}}$ show larger shifts (< 20~$\mathrm{cm^{-1}}$) likely due to larger anharmonic effects.
Several combination bands can be observed, most of which are weak and overshadowed by the strong fundamental bands, except for the three isolated features at 1000, 1246, and 1355 ~$\mathrm{cm^{-1}}$. Another strong feature at 1427~$\mathrm{cm^{-1}}$ is a mix of fundamental and combination bands making the overall feature appear broad. 

We observe a similar doublet feature in the phen$^{2+}$ spectrum at 1333 and 1343~$\mathrm{cm^{-1}}$. As discussed above for naph$^{2+}$, we suspect this blue-shifted feature to be due to a strong combination band involving the Ne-tag. However, similar to naph$^{2+}$, our calculations reveal a Fermi resonance of the fundamental predicted at 1337~$\mathrm{cm^{-1}}$ with two combination modes at 1350 and 1378~$\mathrm{cm^{-1}}$.

\subsection{Anthracene}
\begin{figure*}[htpb]
    \includegraphics[width=18cm]{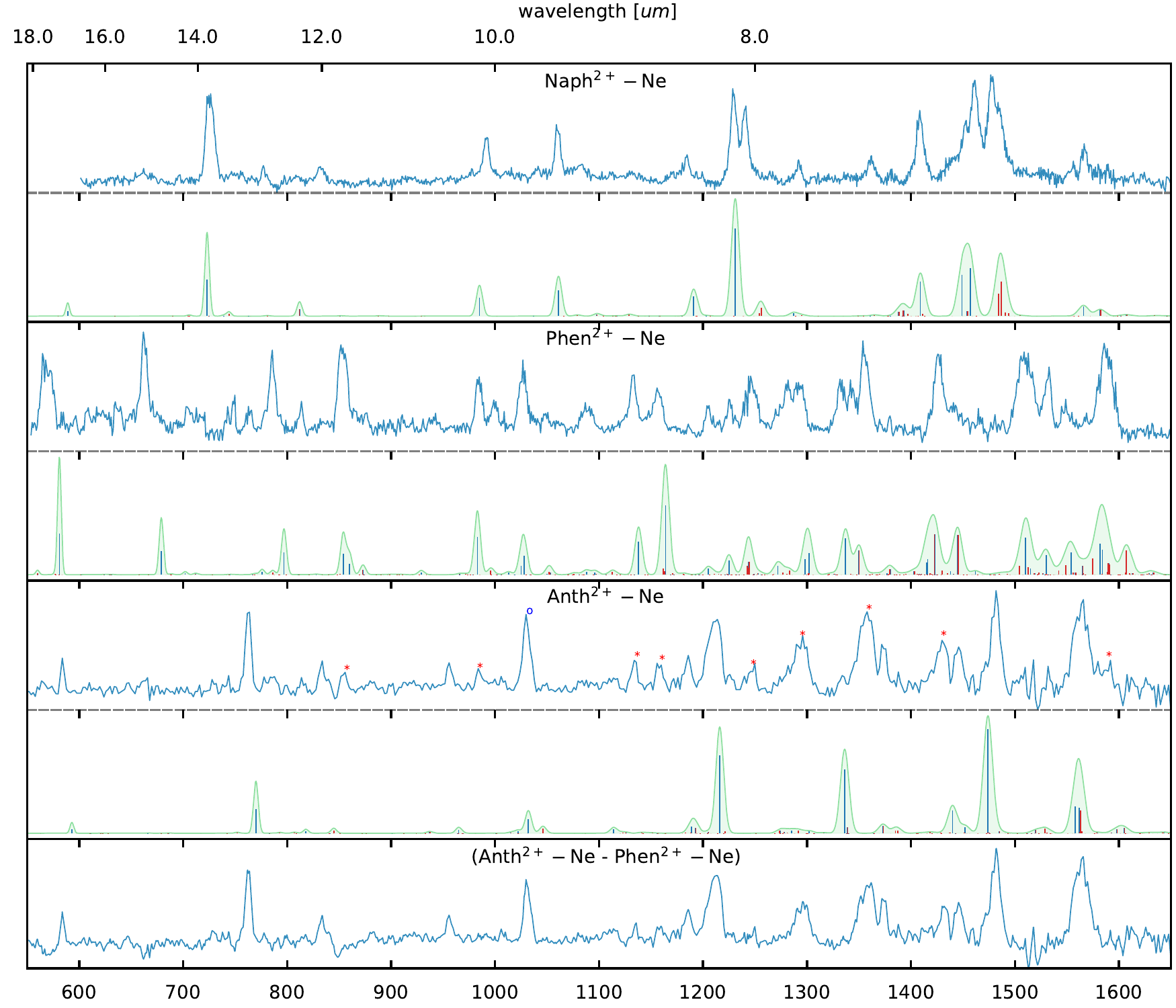}
    \begin{center}
     \large\textsubscript{\textcolor{red}{*} phen$^{2+}$ bands due to contamination,} \hspace{5pt}
    \large\textsubscript{\textcolor{blue}{$^{\circ}$} overlapping phen$^{2+}$ and anth$^{2+}$ bands}
    \end{center}
    \caption{IRPD spectrum of Ne-taggged dications naph$^{2+}$, phen$^{2+}$, and anth$^{2+}$ compared to calculated anharmonic (B3LYP/6-311 G(d,p)) spectra, respectively. The calculated spectrum was convoluted with a Gaussian lineshape function, where the width is given by the FEL bandwidth, and the area corresponds to the calculated intensity in $\mathrm{km\ hspace{1pt}mol ^{-1}}$. The spectrum in the bottom panel is the anth$^{2+}$-Ne spectrum subtracted by 30\% of the weight of the relative intensity of the phen$^{2+}$-Ne spectrum to account for the contaminated bands.}
    \label{All_spec}
\end{figure*}

The lower panels in Fig. \ref{All_spec} show the Ne-IRPD spectrum of anth$^{2+}$ compared to calculated anharmonic spectra of its singlet $^1A_g$ electronic state. The band assignments are shown in Table \ref{Anth_table}. Due to the higher symmetry (D$_{2h}$) of anth$^{2+}$ compared to phen$^{2+}$ (C$_{2v}$), the spectrum shows fewer and less intense vibrational bands. Once again, we do not see any trace of the energetically higher-lying (0.79~eV) triplet $^{3}B_{3g}$ electronic state (\ref{Triplet_table}).

Several bands in the experimental spectrum coincide with the phen$^{2+}$ bands (see caption in Fig. \ref{All_spec}). These phen$^{2+}$ bands appear to be due to contamination from the previous experiment where phenanthrene was introduced in the source.The 1357~$\mathrm{cm^{-1}}$ band, for example, is a convolution of three bands, two of which are assigned to anth$^{2+}$ with some contribution from phen$^{2+}$ contamination of its 1355~$\mathrm{cm^{-1}}$ combination band.

Since phen$^{2+}$ is the more stable isomer based on our calculations (by 0.7~eV), it is possible that the phen$^{2+}$ bands observed in the anth$^{2+}$ spectrum are due to isomerization of anth$^{2+}$ to phen$^{2+}$ during the electron impact ionization process, as has been predicted theoretically for the anthracene monocations \citep{Johansson2011}. To verify this, we also recorded spectra at lower ionization energy (30~eV) expecting a change in the population of the phen$^{2+}$ isomer and consequently a change in the peak intensity of phen$^{2+}$ bands relative to the anth$^{2+}$ bands. This method was  successfully implemented for benzylium and tropylium cations (\ce{C7H7^+}), two isomers formed in dissociative ionization of toluene \citep{Jusko2018a}, but we did not observe any change in the peak intensity for any of the phen$^{2+}$ bands. Instead, the intensity of the phen$^{2+}$ bands decreased over time during the measurement campaign, leading us to conclude that they are due to contamination instead of isomerization. Relative depletion values on the isolated phen$^{2+}$ bands at 786~~$\mathrm{cm^{-1}}$ and 854~$\mathrm{cm^{-1}}$ indicate a contamination of $40-20$\%, decreasing during the measurement campaign. The bottom panel of Fig. \ref{All_spec} shows the difference spectrum where the experimental (scaled) phen$^{2+}$ spectrum was subtracted from the anth$^{2+}$ spectrum to account for this contamination. For this, we used an average 30\% scaling factor over the whole spectral region, whereas the contamination varied from $40-20$~\% over the course of the measurements, as outlined above. This leads to some remaining phen$^{2+}$ artifacts in the difference spectrum, for example, several bands between $1100-1200$~$\mathrm{cm^{-1}}$ have not been completely removed, whereas several negative artifacts appear in the lower wavenumber range. Overall, however, this procedure allows for a better comparison to the calculated spectrum, as given in Table \ref{Anth_table}.

\begin{table*}
    \centering
    \caption[]{Experimentally measured band positions v$_{vib}$  ($\mathrm{cm^{-1}}$), FWHM $\delta$ ($\mathrm{cm^{-1}}$), and relative intensities of the naphthalene dication compared to DFT computed harmonic (scaled by 0.9679) and anharmonic fundamental and combination mode positions and intensities I ($\mathrm{km\hspace{1pt}mol ^{-1}}$) $>$ 5$\mathrm{km\hspace{1pt}mol^{-1}}$. Values in brackets denote uncertainties (1 $\sigma$) in units of the last significant digit.}
    \label{Naph_table}
    \setlength\tabcolsep{4.5pt}
    \begin{tabular}{r c c | r r r| r r r|c}
            \hline \multicolumn{3}{c|}{\text{IRPD (This work)}}
             & \multicolumn{3}{c|}{\text{Anharm. calc.}} & \multicolumn{3}{c|}{\text{Harm. calc.}} & symm
            \\
            \hline
            v$_{vib}$ & $\delta$ & I$_{rel}$  & v$_{vib}$ & I & I$_{rel}$  & v$_{vib}$ & I & I$_{rel}$ 
            \\
            \hline
                                        &               &               &               589     &       10.1    &       0.05    &       577     &       7.7     &       0.03    &       b$_{2u}$        \\
        726(1)  &       9.1     &       0.96    &               723     &       77.2    &       0.42    &       729     &       98.6    &       0.39    &       b$_{3u}$        \\
\tablefootmark{c}       777(1)  &       3.9     &       0.16                            &               &               &               &               &               &               \\
        832(1)  &       6       &       0.12    &       \tablefootmark{a}       812     &       15      &       0.08    &               &               &               &       b$_{2u}$        \\
        992(1)  &       7.7     &       0.45    &               985     &       38.9    &       0.21    &       968     &       33.3    &       0.13    &       b$_{2u}$        \\
        1060(1) &       6.8     &       0.57    &               1061    &       54.1    &       0.29    &       1043    &       67.3    &       0.27    &       b$_{1u}$        \\
        1184(1) &       7.7     &       0.22    &               1191    &       41      &       0.22    &       1172    &       21      &       0.08    &       b$_{2u}$        \\
        1229(1) &       7.3     &       0.96    &               1231    &       184     &       1.00    &       1213    &       231.1   &       0.91    &       b$_{2u}$        \\
        1240(1) &       7.8     &       0.78    &                       &               &               &               &               &               &               \\
\tablefootmark{b}       1266    &               &               &       \tablefootmark{a}       1254    &       6.6     &       0.04    &               &               &               &       b$_{2u}$        \\
                &               &               &       \tablefootmark{a}       1256    &       18.1    &       0.10    &               &               &               &       b$_{2u}$        \\
        1292(1) &       4.9     &       0.16    &               1287    &       6.1     &       0.03    &       1271    &       8.5     &       0.03    &       b$_{1u}$        \\
        1362(1) &       7.8     &       0.20    &               -       &       -       &       -       &               &               &               &               \\
\tablefootmark{b}       1381    &               &               &       \tablefootmark{a}       1388    &       8.9     &       0.05    &               &               &               &       b$_{1u}$        \\
                &               &               &       \tablefootmark{a}       1393    &       12.4    &       0.07    &               &               &               &       b$_{1u}$        \\
        -       &       -       &       -       &       \tablefootmark{a}       1397    &       5.5     &       0.03    &               &               &               &       b$_{1u}$        \\
        1409(1) &       8.9     &       0.67    &               1409    &       72.5    &       0.39    &       1387    &       104.6   &       0.41    &       b$_{1u}$        \\
        1453(2) &       21.5    &       0.49    &               1449    &       86.3    &       0.47    &       1435    &       131.7   &       0.52    &       b$_{1u}$        \\
                &               &               &       \tablefootmark{a}       1454    &       10.6    &       0.06    &               &               &               &       b$_{2u}$        \\
        1461(1) &       7.2     &       0.73    &               1457    &       101.1   &       0.55    &       1451    &       253.7   &       1.00    &       b$_{2u}$        \\
        1480(1) &       17.6    &       1.00    &       \tablefootmark{a}       1484    &       47.4    &       0.26    &               &               &               &       b$_{2u}$        \\
                &               &               &       \tablefootmark{a}       1487    &       73.1    &       0.40    &               &               &               &       b$_{2u}$        \\
                &               &               &       \tablefootmark{a}       1491    &       7.5     &       0.04    &               &               &               &       b$_{1u}$        \\
                &               &               &       \tablefootmark{a}       1494    &       6.9     &       0.04    &               &               &               &       b$_{2u}$        \\
        1567(1) &       21.2    &       0.22    &               1566    &       21.1    &       0.11    &       1553    &       51.8    &       0.20    &       b$_{2u}$        \\
        -       &       -       &       -       &       \tablefootmark{a}       1582    &       13.9    &       0.14    &               &               &               &       b$_{2u}$        \\

            \hline
    \end{tabular}
    \tablefoot{
    \tablefoottext{a}{Combination bands for singlet naph$^{2+}$.}\\
    \tablefoottext{b}{These bands are either too weak or broad due to closely located bands and cannot be fitted.}\\
    \tablefoottext{c}{See text for discussion on this band.}}
\end{table*}

\enlargethispage{1em}
\begin{table*}
    \centering
    \caption[]{Experimentally measured band positions v$_{vib}$  ($\mathrm{cm^{-1}}$), FWHM $\sigma$ ($\mathrm{cm^{-1}}$), and relative intensities of the phenanthrene dication compared to DFT computed harmonic (scaled by 0.9679) and anharmonic fundamental and combination mode positions and intensities I ($\mathrm{km\hspace{1pt}mol ^{-1}}$) $>$ 5$\mathrm{km\hspace{1pt}mol^{-1}}$. Values in brackets denote uncertainties (1 $\sigma$) in units of the last significant digit.}
    \label{Phen_table}
    \setlength\tabcolsep{4.5pt}
    \begin{tabular}{r c c | r r r| r r r}
            \hline \multicolumn{3}{c|}{\text{IRPD (This work)}}
             & \multicolumn{3}{c|}{\text{Anharm. calc.}} & \multicolumn{3}{c}{\text{Harm. calc.}}
            \\
            \hline
            v$_{vib}$ & $\sigma$ & I$_{rel}$  & v$_{vib}$ & I & I$_{rel}$ & v$_{vib}$ & I & I$_{rel}$
            \\
            \hline
                                        569(1)  &       13      &       0.81    &               581     &       78.7    &       0.60    &       569     &       81.3    &       0.34    \\
        663(1)  &       9       &       1       &               679     &       44.9    &       0.34    &       645     &       34.9    &       0.15    \\
        786(1)  &       9       &       0.77    &               797     &       42.3    &       0.32    &       772     &       52.3    &       0.22    \\
        854(1)  &       12      &       0.98    &               854     &       40.1    &       0.30    &       836     &       40.3    &       0.17    \\
                &               &               &               860     &       21.4    &       0.16    &       848     &       35.2    &       0.15    \\
\tablefootmark{b}       875     &               &               &       \tablefootmark{a}       873     &       9.8     &       0.07    &               &               &               \\
        985(1)  &       9       &       0.58    &               983     &       72.7    &       0.55    &       962     &       83.5    &       0.35    \\
        1000(1) &       12      &       0.28    &       \tablefootmark{a}       996     &       7.8     &       0.06    &               &               &               \\
        1027(1) &       10      &       0.65    &               1025    &       16.4    &       0.12    &       1008    &       22.9    &       0.10    \\
                &               &               &               1028    &       36.5    &       0.28    &       1010    &       39.6    &       0.17    \\
        -       &       -       &       -       &       \tablefootmark{a}       1052    &       5.1     &       0.04    &               &               &               \\
        1088(1) &       14      &       0.28    &               1088    &       5.8     &       0.04    &       1069    &       4       &       0.02    \\
        -       &       -       &       -       &       \tablefootmark{a}       1113    &       5.8     &       0.04    &               &               &               \\
        1133(1) &       11      &       0.56    &               1138    &       62.6    &       0.47    &       1120    &       72.9    &       0.31    \\
        1156(1) &       13      &       0.44    &       \tablefootmark{a}       1162    &       11.4    &       0.09    &               &               &               \\
                &               &               &       \tablefootmark{a}       1163    &       6.7     &       0.05    &               &               &               \\
                &               &               &               1164    &       131.9   &       1.00    &       1144    &       154.3   &       0.65    \\
        1205(1) &       7       &       0.28    &               1205    &       11.2    &       0.08    &       1184    &       15.2    &       0.06    \\
        1226(1) &       9       &       0.3     &               1225    &       27      &       0.20    &       1216    &       82.5    &       0.35    \\
        1246(1) &       14      &       0.53    &       \tablefootmark{a}       1243    &       16.6    &       0.13    &               &               &               \\
                &               &               &       \tablefootmark{a}       1244    &       12.9    &       0.10    &               &               &               \\
                &               &               &       \tablefootmark{a}       1244    &       24.5    &       0.19    &               &               &               \\
        1283(2) &       23      &       0.47    &               1272    &       16.9    &       0.13    &       1255    &       29.3    &       0.12    \\
                &               &               &       \tablefootmark{a}       1283    &       7.4     &       0.06    &               &               &               \\
        1294(1) &       8       &       0.49    &               1298    &       30.2    &       0.23    &       1279    &       31.5    &       0.13    \\
                &               &               &               1302    &       41.8    &       0.32    &       1282    &       36.1    &       0.15    \\
        1333(1) &       11      &       0.53    &       \tablefootmark{d}       1337    &       69.9    &       0.53    &       1330    &       235.9   &       1.00    \\
        1343(1) &       6       &       0.47    &       \tablefootmark{a,d}     1350    &       45.9    &       0.35    &               &               &               \\
        1355(1) &       11      &       0.91    &       \tablefootmark{a,d}     1377    &       2.7     &       0.02    &               &               &               \\
                &               &               &               1378    &       0.6     &       0.00    &               &               &               \\
                &               &               &       \tablefootmark{a}       1380    &       10.5    &       0.08    &               &               &               \\
        -       &       -       &       -       &       \tablefootmark{a}       1403    &       6.4     &       0.05    &               &               &               \\
        1427(1) &       13      &       0.79    &               1415    &       23      &       0.17    &       1397    &       4.1     &       0.02    \\
                &               &               &               1416    &       29.3    &       0.22    &       1399    &       164.9   &       0.70    \\
                &               &               &       \tablefootmark{a}       1423    &       77      &       0.58    &               &               &               \\
                &               &               &       \tablefootmark{a}       1430    &       7.5     &       0.06    &               &               &               \\
\tablefootmark{b}       1439    &               &               &               1438    &       7.3     &       0.06    &       1419    &       9.2     &       0.04    \\
                &               &               &       \tablefootmark{a}       1445    &       76.5    &       0.58    &               &               &               \\
        -       &       -       &       -       &               1462    &       6.8     &       0.05    &       1448    &       3.1     &       0.01    \\
        1510(1) &       18      &       0.86    &               1510    &       70.8    &       0.54    &       1493    &       132.3   &       0.56    \\
                &               &               &       \tablefootmark{a}       1504    &       17      &       0.13    &               &               &               \\
                &               &               &       \tablefootmark{a}       1513    &       14.4    &       0.11    &               &               &               \\
                &               &               &               1515    &       11.7    &       0.09    &       1500    &       22.9    &       0.10    \\
        1532(1) &       7       &       0.65    &               1530    &       37.8    &       0.29    &       1519    &       82.3    &       0.35    \\
        -       &       -       &       -       &       \tablefootmark{a}       1542    &       8.2     &       0.06    &               &               &               \\
        1549(4) &       15      &       0.23    &               1554    &       42      &       0.32    &       1540    &       25.2    &       0.11    \\
                &               &               &       \tablefootmark{a}       1549    &       18.3    &       0.14    &               &               &               \\
                &               &               &       \tablefootmark{a}       1565    &       17.2    &       0.13    &               &               &               \\
        1588(1) &       15      &       0.93    &       \tablefootmark{a}       1575    &       30.8    &       0.23    &               &               &               \\
                &               &               &               1582    &       59.5    &       0.45    &       1565    &       200.9   &       0.85    \\
                &               &               &               1584    &       47.7    &       0.36    &       1582    &       100.3   &       0.43    \\
                &               &               &       \tablefootmark{a}       1589    &       5.8     &       0.04    &               &               &               \\
                &               &               &       \tablefootmark{a}       1590    &       22.2    &       0.17    &               &               &               \\
                &               &               &       \tablefootmark{a}       1591    &       19.2    &       0.15    &               &               &               \\
        -       &       -       &       -       &       \tablefootmark{a}       1607    &       46.5    &       0.35    &               &               &               \\

            \hline
    \end{tabular}
    \tablefoot{
    \tablefoottext{a}{Combination bands for singlet naph$^{2+}$. 
    \tablefoottext{b}{These bands are either too weak or broad due to closely located bands and cannot be fitted}. 
    \tablefoottext{c}{See text for discussion on this band.} 
    }
    }
\end{table*}

\begin{table*}
    \centering
    \caption[]{Experimentally measured band positions v$_{vib}$  ($\mathrm{cm^{-1}}$), FWHM $\sigma$ ($\mathrm{cm^{-1}}$) and relative intensities of the anthracene dication compared to DFT computed harmonic (scaled by 0.9679) and anharmonic fundamental and combination mode positions and intensities I ($\mathrm{km\hspace{1pt}mol ^{-1}}$) $>$ 5$\mathrm{km\hspace{1pt}mol^{-1}}$. Values in brackets denote uncertainties (1 $\sigma$) in units of the last significant digit.}
    \label{Anth_table}
    \setlength\tabcolsep{4.5pt}
     \begin{tabular}{r c c | r r r| r r r}
            \hline \multicolumn{3}{c|}{\text{IRPD (This work)}}
             & \multicolumn{3}{c|}{\text{Anharm. calc.}} & \multicolumn{3}{c}{\text{Harm. calc.}}
            \\
            \hline
            v$_{vib}$ & $\sigma$ & I$_{rel}$  & v$_{vib}$ & I & I$_{rel}$ & v$_{vib}$ & I & I$_{rel}$
            \\
            \hline
                        584(1)  &       3.5     &       0.42    &               593     &       13      &       0.038   &       580     &       12.1    &       0.029   \\
                763(1)  &       6.3     &       0.98    &               770     &       80.9    &       0.235   &       757     &       90.9    &       0.219   \\
                834(1)  &       7.1     &       0.29    &               818     &       6.7     &       0.019   &       804     &       12.4    &       0.030   \\
                -       &       -       &       -       &       \tablefootmark{a}       845     &       8.6     &       0.025   &               &               &               \\
        \tablefootmark{c}       855(1)  &       6.6     &       0.19    &               -       &       -       &       -       &       -       &       -       &       -       \\
                956(1)  &       7.7     &       0.29    &               965     &       11.4    &       0.033   &       944     &       13.2    &       0.032   \\
        \tablefootmark{c}       985(3)  &       9.9     &       0.2     &               -       &       -       &       -       &       -       &       -       &       -       \\
                -       &       -       &       -       &               1022    &       6.7     &       0.019   &       992     &       6.8     &       0.016   \\
                1030(1) &       8.8     &       0.82    &               1032    &       47.1    &       0.137   &       1013    &       60.7    &       0.147   \\
                -       &       -       &       -       &       \tablefootmark{a}       1046    &       15      &       0.044   &               &               &               \\
                -       &       -       &       -       &               1114    &       13.4    &       0.039   &       1093    &       21.5    &       0.052   \\
        \tablefootmark{c}       1134(1) &       7.6     &       0.33    &               -       &       -       &       -       &       -       &       -       &       -       \\
        \tablefootmark{c}       1158(1) &       9.2     &       0.29    &               -       &       -       &       -       &       -       &       -       &       -       \\
                1185(1) &       9.3     &       0.35    &               1189    &       21.7    &       0.063   &       1166    &       33.2    &       0.080   \\
                        &               &               &       \tablefootmark{a}       1193    &       18.8    &       0.055   &               &               &               \\
                1211(1) &       16      &       0.81    &       \tablefootmark{a}       1205    &       5.1     &       0.015   &               &               &               \\
                        &               &               &               1216    &       257.4   &       0.748   &       1195    &       292.1   &       0.705   \\
                        &               &               &       \tablefootmark{a}       1221    &       6.1     &       0.018   &               &               &               \\
        \tablefootmark{c}       1246(1) &       11.3    &       0.24    &               -       &       -       &       -       &       -       &       -       &       -       \\
                -       &       -       &       -       &       \tablefootmark{a}       1274    &       8       &       0.023   &               &               &               \\
                -       &       -       &       -       &               1285    &       8.6     &       0.025   &       1263    &       28.1    &       0.068   \\
                -       &       -       &       -       &       \tablefootmark{a}       1292    &       9.2     &       0.027   &               &               &               \\
        \tablefootmark{c}       1293(1) &       18.8    &       0.53    &               -       &       -       &       -       &       -       &       -       &       -       \\
                -       &       -       &       -       &               1302    &       6.1     &       0.018   &       1280    &       6.9     &       0.017   \\
        \tablefootmark{c}       1357(1) &       17.8    &       0.86    &               1336    &       210.7   &       0.612   &       1329    &       280.8   &       0.678   \\
                        &               &               &       \tablefootmark{a}       1339    &       19.5    &       0.057   &               &               &               \\
                1374(1) &       6.3     &       0.47    &       \tablefootmark{a}       1373    &       25.4    &       0.074   &               &               &               \\
                        &               &               &       \tablefootmark{a}       1385    &       9.7     &       0.028   &               &               &               \\
                -       &       -       &       -       &       \tablefootmark{a}       1387    &       8.4     &       0.024   &               &               &               \\
        \tablefootmark{c}       1429(2) &       16.4    &       0.5     &               1440    &       75.5    &       0.219   &       1422    &       101.8   &       0.246   \\
                1446(1) &       8.3     &       0.47    &               1452    &       20.3    &       0.059   &       1435    &       5.3     &       0.013   \\
                1481(1) &       11.3    &       1       &               1474    &       344.1   &       1.000   &       1460    &       414.2   &       1.000   \\
                -       &       -       &       -       &       \tablefootmark{a}       1485    &       7.1     &       0.021   &               &               &               \\
                -       &       -       &       -       &       \tablefootmark{a}       1520    &       10.3    &       0.030   &               &               &               \\
                -       &       -       &       -       &       \tablefootmark{a}       1529    &       15.9    &       0.046   &               &               &               \\
                -       &       -       &       -       &               1558    &       88.3    &       0.257   &       1542    &       114.9   &       0.277   \\
                1564(1) &       20      &       0.89    &               1562    &       85.6    &       0.249   &       1548    &       229.3   &       0.554   \\
                        &               &               &       \tablefootmark{a}       1562    &       5.2     &       0.015   &               &               &               \\
                        &               &               &       \tablefootmark{a}       1563    &       75.1    &       0.218   &               &               &               \\
                        &               &               &       \tablefootmark{a}       1564    &       6.4     &       0.019   &               &               &               \\
                -       &       -       &       -       &       \tablefootmark{a}       1598    &       14      &       0.041   &               &               &               \\
                -       &       -       &       -       &       \tablefootmark{a}       1605    &       18.2    &       0.053   &               &               &               \\

            \hline
    \end{tabular}
    \tablefoot{
    \tablefoottext{a}{Combination bands for singlet naph$^{2+}$.\\
    \tablefoottext{b}{These bands are either too weak or broad due to closely located bands and cannot be fitted.}\\
    \tablefoottext{c}{See text for discussion on this band.} 
    }
    }
\end{table*}

\section{Astrophysical implications and conclusions}
It is known that the charge state of the PAH plays an important role in the IR emission spectrum, especially in the 6-9~$\mu$m region where singly and multiply charged cations of PAH show strong bands \citep{Langhoff1996, Hudgins1999, Allamandola1999, Bauschlicher2000}. \cite{Witt2006} also proposed PAH dications as potential carriers of the extended red emission (ERE) observed in the reflection nebula NGC 7023 and the Red Rectangle nebula via fluorescence. In the ISM, PAHs can reach the doubly charged state in regions of high UV radiation density (e.g., in the PDR of the Orion Bar), through sequential absorption of two photons \citep{Leach1986}. The second ionization energies of PAH cations are below 13.6~eV (the ionization potential of atomic hydrogen), decreasing for larger species \citep{wenzel2020, cccbdb}. In such regions, they play an important role in the chemistry, which in turn is affected by their stability. There has been some discussion in the past on the stability of PAH dications, since their energies are often higher than those of their singly charged fragments due to Coulomb repulsion \citep{Leach1996}. At higher internal energies, 5~eV above the second ionization energy, for example provided during the ionization process, fragmentation channels via covalent dissociation into a smaller dication fragment and a neutral also open up. We see this channel in the mass spectrum for anthracene electron impact ionization above energies of 23(2)~eV with the appearance of a mass peak at m/z=76, which we interpret as the \ce{C12H8} fragment dication. However, often the dissociation is hindered by barriers in the potential energy surface along the dissociation coordinate, as has been shown in the case of the benzene dication \citep{ROSI2004, Jasik2014}. Earlier mass-spectrometric studies reported a yield of roughly 10~\% for PAH dications upon ionization of the neutral, with only marginal variations for different PAH sizes and ionization methods (e.g., EI and photo-ionization) \citep{Leach1989a,Leach1989b,Zhen2017,Zhen2016}. We see a similar behavior for the three PAH dications considered in this study. As exemplarily shown for the anthracene ions in Figs. \ref{ionyield} and \ref{massspec}, we reach dication to monocation ratios of up to 30~\% at high electron impact energies, and a dication yield of around 10~\% compared to all observed fragment plus parent ions. Evaluation of the abundance of PAH dications in the ISM, where they need to be produced by sequential photo-ionization competing with dissociation, chemical reactions, and recombination processes, requires detailed modeling of the PAH evolution. In particular, the comparably small PAH dications studied here are likely not stable in interstellar conditions \citep{Montillaud2013, Zhen2015, Zhen2016}. It would therefore be interesting to extend the vibrational studies presented here to the class of larger, astronomically more relevant PAH dications.

To date, vibrational spectral information for PAH dications comes mainly from theoretical calculations, and we are aware of only one experimental IR study by \citet{Zhen2018}. \citeauthor{Malloci2007} did a comparative theoretical study on IR properties of 40 different PAH neutrals, monocations, and dications using DFT and TD-DFT theoretical techniques. Based on these calculations, most of the dications were predicted to have a singlet ground state including the dications presented in this work and our recorded spectra support this finding. Another notable characteristic in the recorded IR spectra of all three dications is the presence of several intense combination bands, especially in the 5$-$10~$\mu$m region. Here, anharmonic calculations proved to be crucial in assigning these combination bands. The presence of combination bands has an effect on the fraction of the total integrated intensity (InI) in the different spectral ranges 2.5-3.5, 5-10, 10-15~$\mu$m, and >15~$\mu$m. Differences in the relative integrated intensities in these spectral ranges serve as identifiers for the charge state of PAHs in the ISM, as has been discussed in detail previously \citep{Bauschlicher2000, Malloci2007}. Whereas drastic changes in the integrated intensities were observed between neutral and cationic species, there were much fewer predicted variations between the singly and doubly charged PAHs. The question arises if the appearance of strong combination bands changes this picture.      
A comparison is made in Appendix \ref{sec:DFT Analysis} showing the effect of including anharmonicity in the calculations on the integrated intensities of singly and doubly charged PAHs considered in this work. As the choice of the basis set in theoretical calculations of vibrational spectra has been discussed previously \citep{Andersson2005,Bauschlicher1997}, we first compared the InI for the two functionals B3LYP/6-311G (d, p) (this work) and B3LYP/4-31G \citep{Malloci2007}, using the harmonic approximation, see Fig. \ref{InI2} in Appendix \ref{sec:DFT Analysis}. We can see that there is no significant effect on InI for either the singly or doubly charged species. However, when performing anharmonic calculations (with the B3LYP/6-311G (d, p) functional), the InI in the 2.5-5~$\mu$m region is 5-7\% higher for both singly and doubly charged PAHs compared to harmonic calculations using the same functional. In contrast, in the 5-10$\mu$m region, InI predicted by anharmonic calculations is lower compared to harmonic calculations by 28~\% for naph$^{+1}$, 15\% for anth$^{+1}$, and phen$^{+1}$, but only 2-5~\% for doubly charged PAHs.  In the spectral regions >15~$\mu$m and 10-15~$\mu$m, both singly and doubly charged anth and phen show negligible (<1~\%) changes in InI, whereas naph shows 7~\% and 12~\% higher InI predicted with anharmonic compared to harmonic calculations. Taking into account the fact that the absolute intensities of the fundamental bands are generally predicted lower in anharmonic calculations than harmonic (see tables \ref{Phen_table}, \ref{Anth_table}), about 45~\% and 25~\% of the absolute integrated intensity arises entirely from combination bands of singly and doubly charged phen and anth, respectively.

The significant differences observed in the relative integrated intensities in each of the spectral regions and the presence of combination bands suggest that anharmonic calculations are necessary to predict accurate vibrational spectra for the family of comparatively small PAH cations as discussed here and also in previous work \citep{Maltseva2015, Lemmens2019}. Thus, including these anharmonic effects has consequences on the interpretation of relative UIR band intensities observed in the ISM.

Here, we have shown that cryogenic IRPD experiments using Ne-tagging prove to be a very powerful method to obtain vibrational spectra of PAH dications by providing narrow features and intensities being more comparable to the linear absorption cross section. This makes it a very effective tool to benchmark theoretical calculations. In order to validate our finding that anharmonic effects significantly influence the IR band positions and intensities of PAH monocations and dications, future IRPD experiments targeting larger and structurally different classes of PAH cations should be conducted. As discussed above, the double ionization process competes with fragmentation channels. One of the main fragmentation channels observed for PAHs is the loss of acetylene, \ce{C2H2} \citep{Johansson2011,Simon2017,Ling1998}, as we have also observed in this study (see Fig. \ref{massspec}). Elucidating the structure of these (\ce{C2H2})-loss fragment ions, as has been previously done for the case of naphthalene \citep{Bouwman2016}, is another interesting application for the narrow-linewidth IRPD action spectroscopy that we have presented here. 

\begin{acknowledgements}
This project is funded by the Marie Skłodowska Curie Actions (MSCA) Innovative Training Networks (ITN) H2020-MSCA-ITN 2016 (EUROPAH project, G. A. 722346). We are grateful for the experimental support provided by the FELIX team and acknowledge the Nederlandse Organisatie voor Wetenschappelijk Onderzoek (NWO) for the support of the FELIX Laboratory. we also thank NWO Exact and Natural Sciences for the use of supercomputer facilities (Grant nr. 2019.062). We thank the Cologne Laboratory Astrophysics group for providing the FELion ion trap instrument for the current experiments and the Cologne Center for Terahertz Spectroscopy (core facility, DFG grant SCHL 341/15-1) for supporting its operation. We would like to thank the anonymous referee for valuable comments which helped to improve the manuscript.
\end{acknowledgements}

\bibliographystyle{aa}
\bibliography{mybib}

\begin{appendix}
\onecolumn

\section{Mass spectrum showing doubly charged anthracene tagged with Ne}
\begin{figure*}[htpb]
\includegraphics[width=18cm]{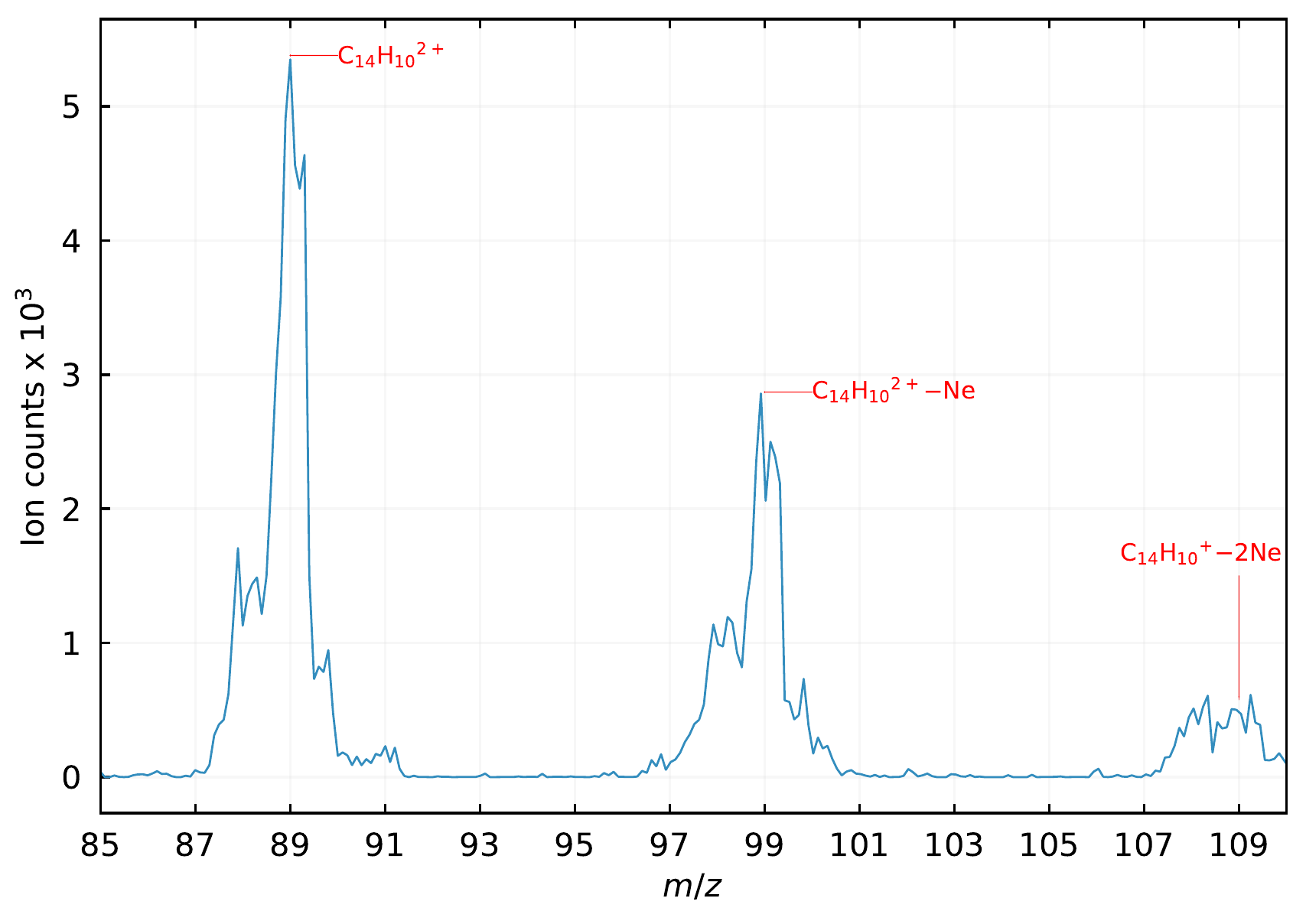}

  \caption{Mass spectrum showing anth$^{2+}$ (\ce{C14H10$^{2+}$}, m/z=89) with one (\ce{C14H10$^{2+}$-Ne}, m/z=99) and two  (\ce{C14H10$^{2+}$}-Ne$_2$, m/z=109) attached neon atoms. Additional mass peaks appearing with $\Delta$m/z=0.5 can be seen in the spectrum for the bare ion and Ne-dication complexes. They belong to dehydrogenated phen$^{2+}$ (at lower mass) and $^{13}$C substituted or hydrogenated phen$^{2+}$ (higher masses), respectively, see also \citet{vanderBurgt2018}. The mass resolution of the second quadrupole mass filter is better than m/z=0.5, which allowed us to mass select only the singly tagged target complex to record the IRPD spectra.}
     \label{mass_spec_Ne_tagging}
\end{figure*}

\pagebreak

\section{Comparison of vibrational spectra of bare and Ne-tagged Naph$^{2+}$}

\begin{figure*}[htpb]
  \centering
\includegraphics[width=18cm]{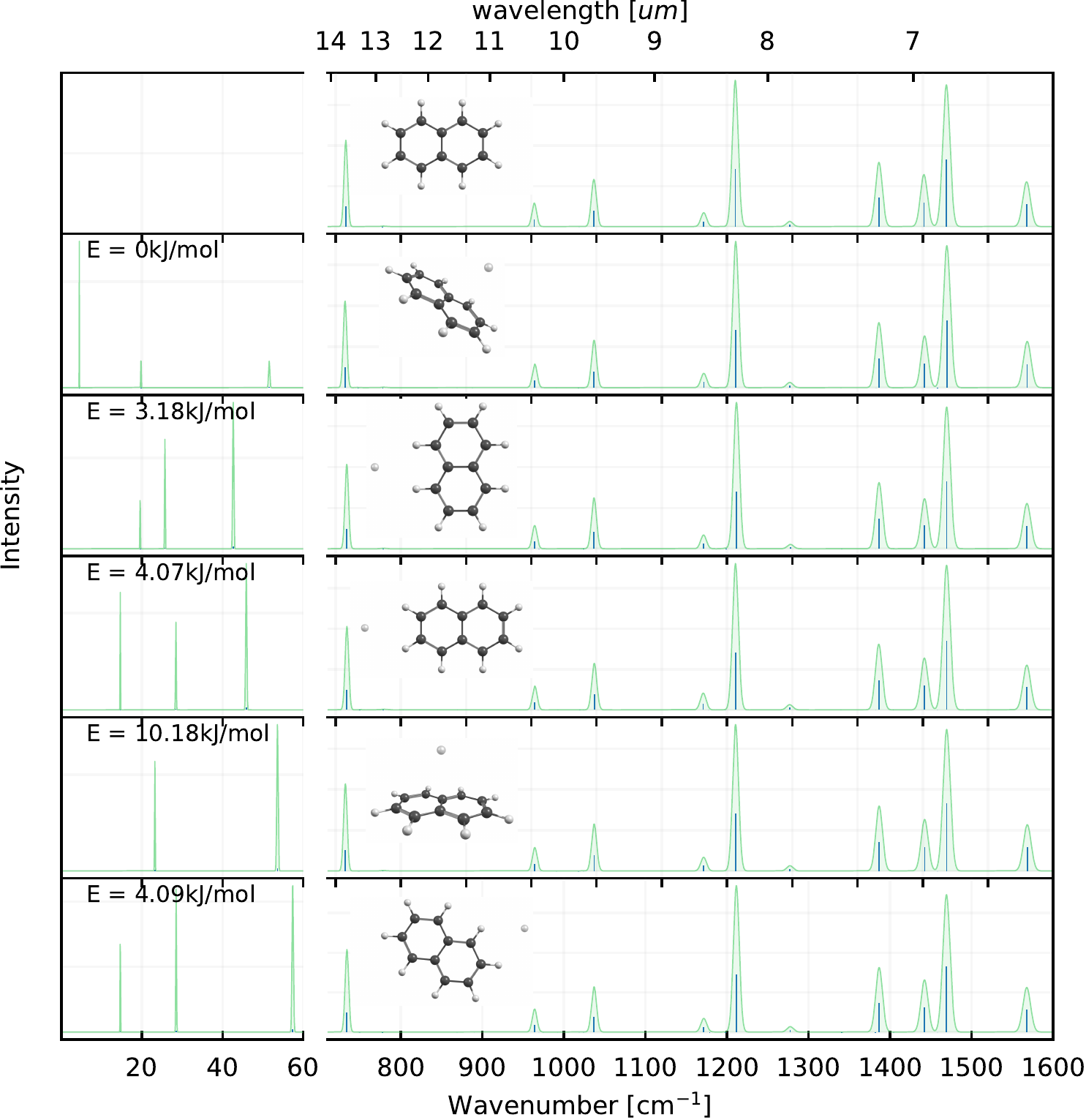}
\caption{Vibrational spectra of doubly charged bare naphthalene and neon-tagged naphthalene calculated using DFT within the harmonic approximation at the  wB97XD/cc-pVTZ level. The respective energies of the five lowest conformers relative to the lowest energy conformer are stated. The frequency shifts between the different neon isomers and the bare ion are negligible. The three bands observed in the 0-60~cm$^{-1}$ region are the bending and stretching vibrations involving the neon atom where we observe large shifts in frequencies among its isomers.}
\label{Ne_tagging}
\end{figure*}
\break

\section{Comparison between anharmonic and harmonic calculations}\label{sec:DFT Analysis}

\begin{figure*}[htpb]
\includegraphics[width=18cm]{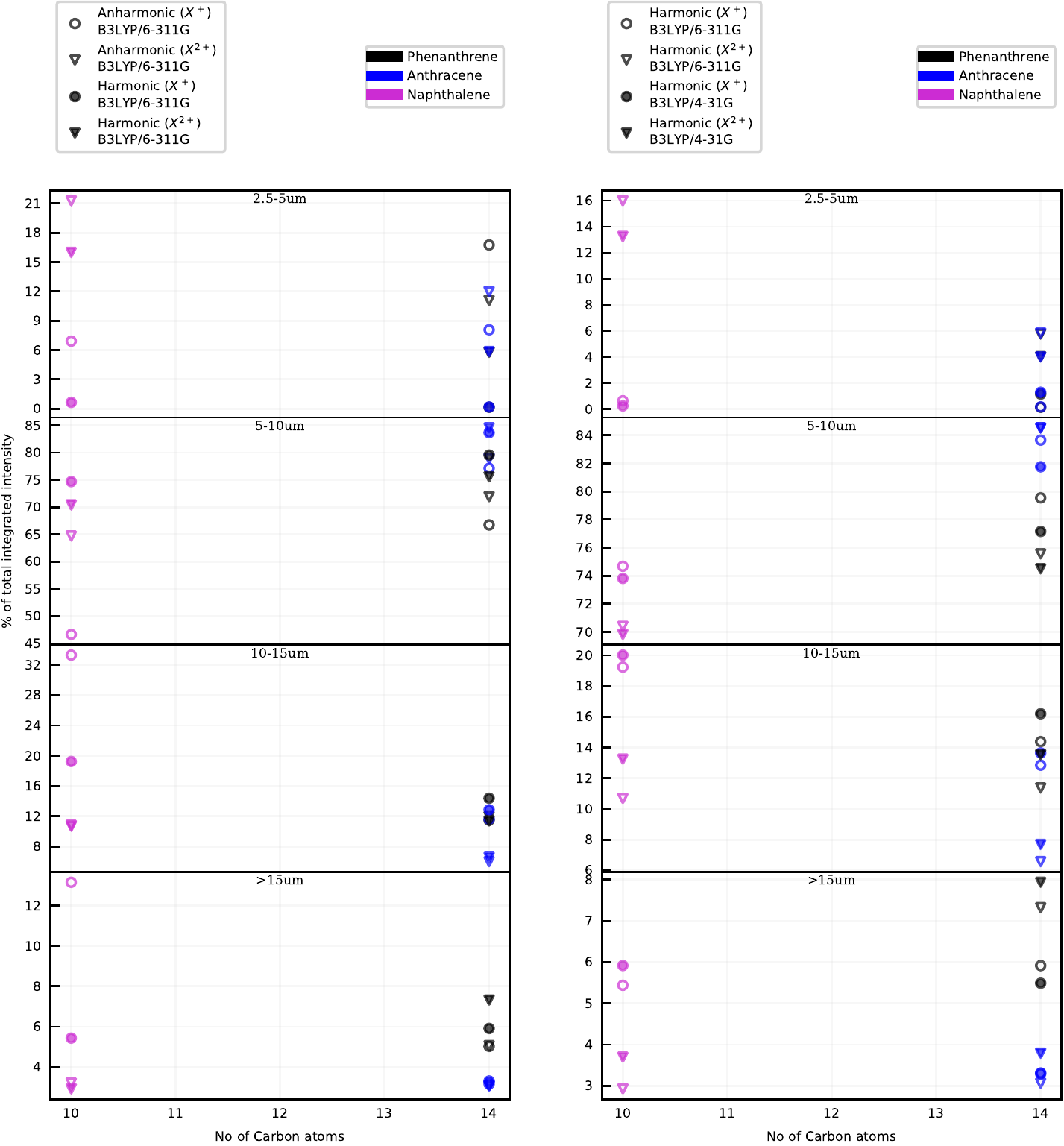}
  \caption{Integrated intensities in different regions of the mid IR spectra. Left Panel: Comparison made between harmonic calculations using the B3LYP/6-311-G(d,p) and  B3LYP/4-31G functionals. Right panel: Comparison made between anharmonic and harmonic calculations using the  B3LYP/6-311-G(d,p) functional.}
     \label{InI2}
\end{figure*}

\begin{table*}
    \centering
    \caption[]{DFT computed harmonic (scaled by 0.9679) fundamental mode positions $v_{vib}$ ($\mathrm{cm^{-1}}$) and intensities I ($\mathrm{km\hspace{1pt}mol ^{-1}}$) > 5$\mathrm{km\hspace{1pt}mol ^{-1}}$ of triplet electronic ground state for naph$^{2+}$, anth$^{2+}$, and phen$^{2+}$.}
    \label{Triplet_table}
    $
    \begin{tabular}{ r r | r r | r r  }
        \hline
        \multicolumn{2}{c|}{\text{Naph$^{2+}$}}
        &  \multicolumn{2}{c|}{\text{Anth$^{2+}$}} &    \multicolumn{2}{c}{\text{Phen$^{2+}$}}\\
        \hline
         v$_{vib}$ & I &   v$_{vib}$ & I & v$_{vib}$ & I \\
        \hline
        137     &       5.1     &       252     &       28.2    &       346     &       12      \\
407     &       17.6    &       410     &       26.7    &       401     &       8.1     \\
755     &       97.5    &       548     &       23.9    &       712     &       91.8    \\
922     &       216.9   &       715     &       87.6    &       787     &       22      \\
976     &       9.6     &       741     &       113.6   &       808     &       47.1    \\
1064    &       57.4    &       781     &       153.8   &       849     &       13.7    \\
1138    &       32.6    &       864     &       50.9    &       927     &       119.9   \\
1186    &       15.2    &       880     &       31.9    &       991     &       9.6     \\
1189    &       11.8    &       902     &       296.9   &       1019    &       20.6    \\
1365    &       19.7    &       943     &       16.5    &       1046    &       9.4     \\
1401    &       269.2   &       1133    &       13.1    &       1085    &       10.5    \\
1416    &       250.9   &       1294    &       210.9   &       1114    &       18      \\
3082    &       8       &       1349    &       386.7   &       1164    &       45      \\
3089    &       94.7    &       1373    &       81.8    &       1170    &       168.5   \\
3093    &       91.6    &       1376    &       269.5   &       1209    &       6       \\
        &               &       1393    &       42.5    &       1302    &       7.8     \\
        &               &       2982    &       7.5     &       1310    &       24.1    \\
        &               &       2989    &       6.1     &       1364    &       8.8     \\
        &               &       2998    &       41.8    &       1381    &       101.4   \\
        &               &       3003    &       48.2    &       1385    &       34.2    \\
        &               &               &               &       1422    &       59.4    \\
        &               &               &               &       1439    &       13.8    \\
        &               &               &               &       1441    &       87.2    \\
        &               &               &               &       1490    &       44.3    \\
        &               &               &               &       1509    &       25.6    \\
        &               &               &               &       2996    &       32.2    \\
        &               &               &               &       3002    &       49      \\
        &               &               &               &       3009    &       11.2    \\
        &               &               &               &       3022    &       10.4    \\

        \hline
    \end{tabular}
    $
\end{table*}

\begin{figure*}
    \begin{center}
    \includegraphics[width=6cm]{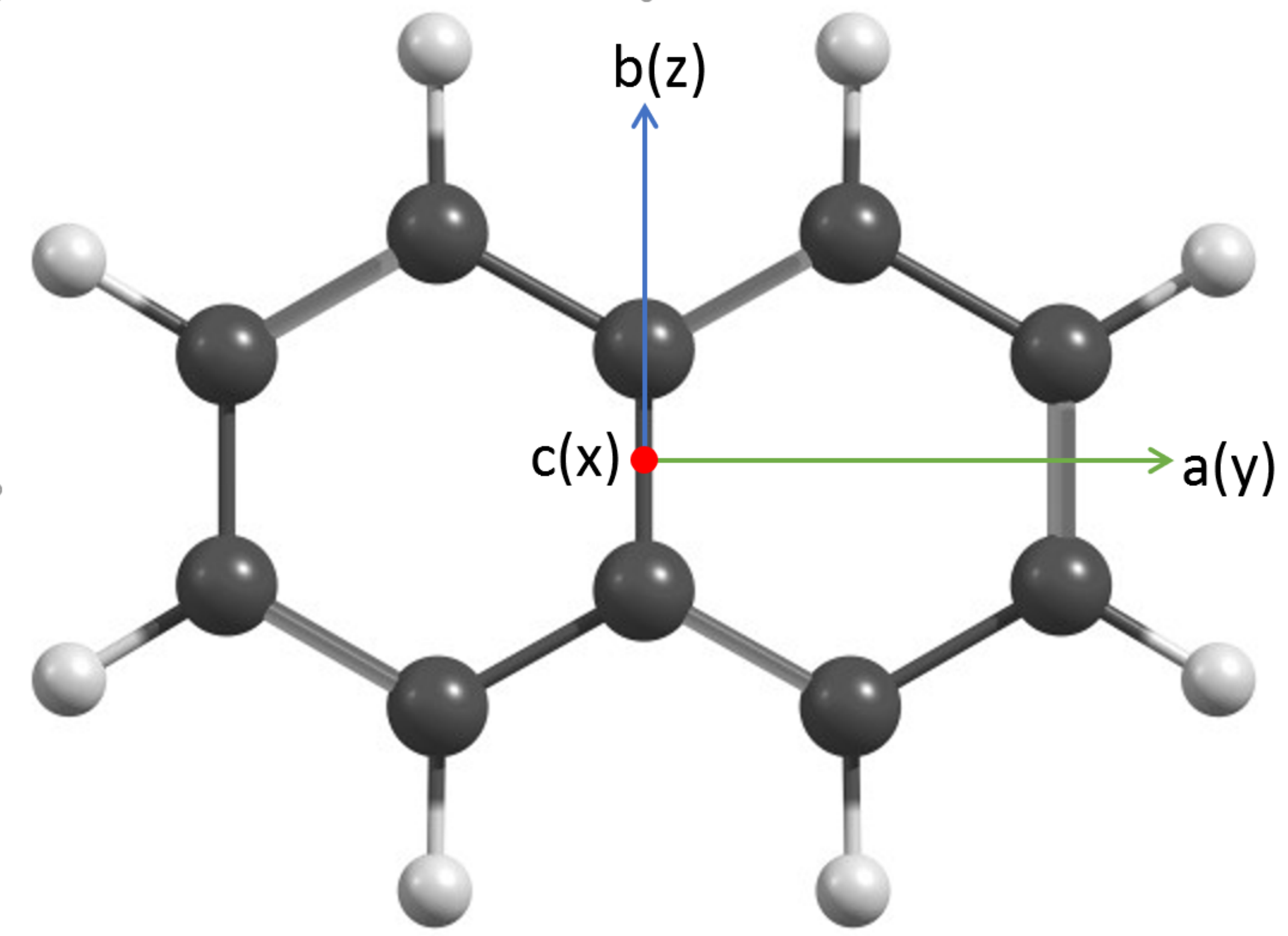}
    \caption{Naphthalene dication showing the orientation of the molecular axes used in the calculations ($II^r$ representation), defining the symmetry labels given in Table \ref{Naph_table}. Inertia moments: Iy < Iz < Ix.}
    \end{center}
\end{figure*}
\end{appendix}
\end{document}